\documentclass{jfm}
\usepackage{bm}
\usepackage{graphicx}
\usepackage{epstopdf, epsfig}
\usepackage{color} 
\usepackage{amsmath}

\usepackage{cases}

\shorttitle{Drag Coefficient of a Rigid Spherical Particle in a Binary Fluid Mixture}
\shortauthor{S. Yabunaka and Y. Fujitani}

\title{Drag Coefficient of a Rigid Spherical Particle \\
in a Near-Critical Binary Fluid Mixture \\
beyond the Regime of the Gaussian Model}

\author{Shunsuke Yabunaka}
\affiliation{
Department of Physics, Kyushu University, Fukuoka 819-0395, Japan
}

\author{Youhei Fujitani}
\affiliation{
School of Fundamental Science and Technology, Keio University, Yokohama 223-8522, Japan 
}

\author{Shunsuke Yabunaka\aff{1}
  \corresp{\email{yabunaka123@gmail.com}},
 \and Youhei Fujitani\aff{2}
  \corresp{\email{youhei@appi.keio.ac.jp}}}

\affiliation{\aff{1}Department of Physics, Kyushu University, Fukuoka 819-0395, Japan
\aff{2}School of Fundamental Science and Technology, Keio University, Yokohama 223-8522, Japan}

\begin{document}

\maketitle

\begin{abstract}
The drag coefficient of a rigid spherical particle deviates from the Stokes law
when it is put into a near-critical fluid mixture in the homogeneous phase with the critical composition.
The deviation ($\Delta\gamma_{\rm d}$) is experimentally shown to depend approximately linearly on 
the correlation length \textcolor{black}{of the composition fluctuation} far from the particle ($\xi_\infty$), and is suggested to be caused by
the preferential  \textcolor{black}{adsorption} between one component and the particle surface.   In contrast,
the dependence was shown to be much steeper in the previous theoretical studies
based on the Gaussian free-energy density.  In the vicinity of the particle,
especially when the adsorption of the preferred component makes the composition strongly off-critical, 
the correlation length becomes very small as compared with $\xi_\infty$.
This spacial inhomogeneity, not considered in the previous theoretical studies, 
can influence the dependence of $\Delta\gamma_{\rm d}$ on $\xi_\infty$. 
 To examine this possibility,
we here apply the \textcolor{black}{ renormalized local }functional theory, \textcolor{black}{describing} 
the preferential adsorption in terms of the surface field.  
\textcolor{black}{This} theory was previously proposed to explain the interaction of walls
immersed in a (near-)critical binary fluid mixture.
The free-energy density in this theory, coarse-grained up to the local correlation length, has much complicated dependence
on the order parameter, as compared with the Gaussian free-energy  density.  Still, a concise expression of the drag coefficient, 
which was
derived in one of the previous theoretical studies, turns out to be available in the present formulation.
We show that, as $\xi_{\infty}$ becomes larger, 
the dependence of $\Delta\gamma_{\rm d}$ on $\xi_\infty$  
becomes distinctly gradual and close to the linear
dependence.\end{abstract}

\section{Introduction\label{sec:intro}}
\textcolor{black}{B}rownian motion has been one of the main topics in the physics for a long time \citep{Brev}.
When a rigid spherical particle moves translationally at a sufficiently low constant speed in a quiescent one-component  fluid, the fluid exerts a drag force,
whose magnitude is proportional to the particle speed.  The constant of
proportionality, called the drag coefficient,
is given by $6\pi \eta_{\rm o} a$, where $\eta_{\rm o}$ is the fluid viscosity and $a$ is the particle radius, 
according to the Stokes law \citep{stokes}.  The self-diffusion coefficient of a colloidal particle 
is usually equal to $k_B T$ divided
by the drag coefficient\citep{suther, eins}, where
$k_B$ is the Boltzmann constant and $T$ is the temperature of the fluid. 
This relation (Einstein's relation) can be derived from a linear Langevin equation for the particle velocity
even if it is generalized to have the memory kernel
representing the back-flow effect \citep{zwan,zwan2,widom,case,kubo}.
A Brownian particle has been
used experimentally as a probe 
for local environments in the field of microrheology \citep{passive, microrh, kimura, cell, viscel, viscel2}.
The trajectory of an optically trapped Brownian particle can be measured with resolutions of nanometers and microseconds \citep{lukic,nat,natp,grim}.  

Some properties of a binary fluid mixture 
in the homogeneous phase near the demixing critical point can be also probed by a Brownian particle in the mixture \citep{Bonn, Mazur,ZwanMazur,furu,araki,camley,newraft}. 
\textcolor{black}{Although it is not always the case} \textcolor{black}{\citep{beysens2019}},
the particle surface usually attracts one of the two components more; this
preferential  \textcolor{black}{adsorption}
can be described using a surface field in a coarse-grained picture.
As a result, the preferred component is
absorbed near the particle surface; the composition deviation
from the composition in the bulk
decays very slowly in the adsorption layer, which extends from the surface by the thickness comparable 
with the correlation length of the \textcolor{black}{composition fluctuation} in the bulk
\citep{Cahn, binder, Beysens, beysens1985, holyst}. 
It has been observed that the self-diffusion coefficient of a particle, put in a mixture with the critical composition, 
becomes smaller as the critical temperature is approached on the side of the homogeneous phase 
\citep{Martynets, Martynets2, Lyons, Lyons2, Lee, Omari}. 
Let us write $\gamma_{\rm d}$, $\Delta\gamma_{\rm d}\equiv\gamma_{\rm d}-6\pi\eta_{\rm o}a$, and $\xi_\infty$
for the drag coefficient, the deviation of $\gamma_{\rm d}$ 
from the Stokes law, and the correlation length far from the particle, respectively. 
Some researchers \citep{Lee,Omari}
interpret their data suggesting the linear dependence of 
$\Delta\gamma_{\rm d}$ on $\xi_\infty$ by assuming the particle radius to be effectively 
enlarged by the thickness of the adsorption layer, which is on the order of  $\xi_\infty$. 
Clearly, \textcolor{black}{however}, the adsorption layer is not a part of a rigid particle; it can be deformed and is open to fluid flows. 
It is thus necessary to explain the deviation in terms of the hydrodynamics which can describe the flow in the adsorption layer.
In this line of study,  Okamoto ${\it et\, al.}$ \citep{Okamoto-Fujitani-Komura} employed
the hydrodynamics based on the Gaussian free-energy density
with the surface field being considered, 
and revealed that the osmotic pressure due to the composition gradient around the particle can cause the 
deviation. However, their result of $\Delta\gamma_{\rm d}$ 
is proportional to $\xi_\infty^6$, which is
much steeper than the observed dependence. Their calculation,
supposing sufficiently weak adsorption, was extended to treat  strong adsorption in the framework based on the Gaussian free-energy density
 in \citep{Fujitani2018}, 
where the dependence of $\Delta\gamma_{\rm d}$ on $\xi_\infty$ is still shown to be steeper than linear.
The Gaussian model used in these previous studies supposes a small and homogeneous correlation length.  
However, as the adsorption is stronger,
the correlation length \textcolor{black}{of the fluctuation} near the particle becomes smaller than $\xi_{\infty}$ because the 
\textcolor{black}{average} composition 
there becomes farther from the critical composition, which 
is realized far from the particle
\citep{OkamotoOnuki}.   
This inhomogeneity can reduce  
the dependence of $\Delta\gamma_{\rm d}$ on $\xi_\infty$ from the one
in the Gaussian model.  

\textcolor{black}{After giving a brief review on
the conventional Stokes law in Sect.~\ref{sec:prel},}
we apply the renormalized local functional theory 
to consider the inhomogeneity \textcolor{black}{in a near-critical binary fluid mixture},
instead of the Gaussian model \textcolor{black}{used previously.}
This theory was originally proposed by Fisher and Au-Yang \citep{FisherAu-Yang} 
for static properties of binary fluid mixtures at
the critical point and extended by Okamoto and Onuki \citep{OkamotoOnuki} to describe 
those near the critical point.  In this theory,
the free-energy density of the Ginzburg-Landau-Wilson type  
is coarse-grained up to the  local \textcolor{black}{fluctuation} correlation length,
as mentioned in Sect.~\ref{sec:statics}. 
From this density, the hydrodynamics for the length scales larger than the  local \textcolor{black}{correlation length}
can be formulated
\citep{Yabu-Oka-On-Soft}, which is mentioned in 
Sect.~\ref{sec:dyn}.  The procedure of formulating the hydrodynamics is the same as in the model H (a standard model for the critical dynamics)\textcolor{black}{\citep{Hohen-Halp, Onuki-1}};
the thermal noise need not be considered here for the dynamics at the large length scales 
\citep{Okamoto-Fujitani-Komura, furu}.
\textcolor{black}{In fact,} considering the thermal noise in a near-critical mixture
does not change the Stokes law \textcolor{black}{in the absence of} the preferential  \textcolor{black}{adsorption}, according to
\citep{Mazur}.
The Gaussian free-energy density is composed of a quadratic function with respect to the order parameter and the square gradient term, 
while the density is much more complicated in the renormalized local functional theory.
Still, a concise expression of the drag coefficient, which was previously
derived in the Gaussian model \citep{Fujitani2018},
is shown in Sect.~\ref{sec:perturb} to remain available in the present study.
The perturbation with respect to a dimensionless surface field,
employed in \citep{Fujitani2018}, is not available here.  We instead devise an alternative procedure in Sect.~\ref{sec:calQ}. 
Two points peculiar to our calculation are
stated in the rest of Sect.~\ref{sec:calc}.
Our results are shown in Sect.~\ref{sec:res}, and are
discussed in Sect.~\ref{sec:discuss}. 
\textcolor{black}{\section{Preliminaries\label{sec:prel}}
In this section,} \textcolor{black}{we assume that a rigid spherical particle
undergoes translational motion with a constant velocity in a quiescent incompressible fluid
which can be regarded as a one-component fluid, to
review a} \textcolor{black}{hydrodynamic 
derivation of the conventional} \textcolor{black}{Stokes law. 
The derivation below} \textcolor{black}{can} 
\textcolor{black}{be extended straightforwardly for our problem discussed in the subsequent sections.}  \textcolor{black}{
See \citep{bedeaux} for
the derivation from the linearized equations in the fluctuating hydrodynamics, and \citep{itami}
for the one from the Hamilton dynamics of many particle systems 
within the linear response regime.} 

\textcolor{black}{
The space and time variables are denoted by $({\bm x},t)$;
$\rho_{\rm o}$ and $\eta_{\rm o}$ respectively denote the mass density  and the viscosity of the ambient fluid.
  Its velocity and pressure fields, 
${\bm v}$ and $p$,
satisfy the incompressibility condition and the Navier-Stokes equation, {\it i.e.\/},
\begin{equation}
0=\boldsymbol{\textcolor{black}{\boldsymbol{\nabla}}} \cdot {\bm v}\quad {\rm and}\quad \rho_{\rm o}\frac{D{\bm v}}{Dt}=-\textcolor{black}{\boldsymbol{\textcolor{black}{\boldsymbol{\nabla}}}} p+\eta_{\rm o}\Delta {\bm v}
\ ,\label{eqn:onecomp}\end{equation}
where $D/(Dt)$ implies the Lagrangian time-derivative.
The no-slip boundary condition is imposed at the particle} \textcolor{black}{surface.}

\textcolor{black}{We write $a$ and $U_{\rm o}{\bm e}_z$ for the particle radius and the particle velocity, respectively,
where ${\bm e}_z$ denotes a unit vector.  The fields are stationary if viewed 
on the frame co-moving with the partcle, where the space and time variables} \textcolor{black}{are}  \textcolor{black}{denoted by
$({\bm x}', t')$.  Because of ${\bm x}'={\bm x}-U_{\rm o}t{\bm e}_z$
and $t'=t$, the second entry of (\ref{eqn:onecomp}) becomes
\begin{equation}
\rho{\bm v}'\cdot\textcolor{black}{\boldsymbol{\textcolor{black}{\boldsymbol{\nabla}}}}'{\bm v}'=-\textcolor{black}{\boldsymbol{\textcolor{black}{\boldsymbol{\nabla}}}}'p'+\eta_{\rm o}\Delta' {\bm v}'
\ ,\label{eqn:comove}\end{equation}
where ${\bm v}'={\bm v}-U_{\rm o}{\bm e}_z$ and $p'=p$ are the fields on the co-moving frame, and
$\textcolor{black}{\boldsymbol{\textcolor{black}{\boldsymbol{\nabla}}}}'$ and $\Delta'$ are the differential operators with respect to ${\bm x}'$. 
We can non-dimensionalize the fields by using $\rho_{\rm o}$, $a$ and $U_{\rm o}$ to have the
Reynolds number.  Assuming the low Reynolds number \citep{lowreynolds,Hutter}, we neglect the
left-hand side of (\ref{eqn:comove}), and thus that of the second entry of} \textcolor{black}{ 
(\ref{eqn:onecomp}).
The resultant equation is the Stokes equation} \textcolor{black}{on the original frame.
Linearizing the governing equations in this way, we} \textcolor{black}{
obtain} \textcolor{black}{${\bm v}$ and $p$
up to the linear order of the particle velocity to calculate the drag coefficient.}

\textcolor{black}{The linearization above is also attainable without} \textcolor{black}{explicit non-dimensionalization} 
\textcolor{black}{if we use $\varepsilon U$ instead of $U_{\rm o}$,   
where $U$ is a nonzero constant with the dimension of speed and
$\varepsilon$ is a dimensionless parameter.}   \textcolor{black}{ What to do 
for the linearization with respect to the particle speed is} \textcolor{black}{ 
to expand ${\bm v}$ and $p$ with respect to $\varepsilon$
and perform calculations only up to the order of $\varepsilon$.} \textcolor{black}{Although $\varepsilon$
is known to represent the Reynolds number in this example,  we need not specify
this physical meaning for calculating }\textcolor{black}{the drag coefficient, which is the ratio of
the magnitude of the drag force to the particle speed in the limit of $\varepsilon\to 0$.  
We define ${\bm v}^{(1)}$ and $p^{(1)}$ so that}
\begin{equation}
{\bm v}=\varepsilon {\bm v}^{(1)}\quad {\rm and}\quad p=p^{(0)}+\varepsilon p^{(1)}
\label{eqn:vpexp}\end{equation}
\textcolor{black}{hold up to the order of $\varepsilon$, where the static pressure $p^{(0)}$ is a constant.}
\textcolor{black}{Using the co-moving frame transiently, as in the preceding paragraph,
we have the incompressibility condition and Stokes equation for ${\bm v}^{(1)}$ and $p^{(1)}$.
The solution in \citep{Lamb} is available as it is in calculating the drag coefficient, 
but we here \textcolor{black}{show} an alternative way to prepare for the
calculations in the subsequent sections.}  

We set the spherical polar coordinate system $(r, \theta, \phi)$ so that
\textcolor{black}{the polar axis ($z$-axis) is along ${\bm e}_z$, }and consider the instance that the particle center passes the origin.
Because of the symmetry of the particle motion, we use a spherical harmonics 
$Y_{10}\left(\theta\right)=\sqrt{3/\left(4\pi\right)}\cos\theta$ to \textcolor{black}{express the angular dependence of $p^{(1)}$
\textcolor{black}{as
\begin{equation}
p^{(1)}({\bm x})=p_{10}(r)Y_{10}(\theta)\ ,\label{eqn:defp10}
\end{equation}
whereby $p_{10}$ is defined.}
Using the vector spherical harmonics for that of ${\bm v}^{(1)}$ \citep{Fujitani2007},
we define $R_{10}$ and $T_{10}$ so that  }
\begin{equation}
v_r^{(1)}({\bm x})=R_{10}\left(r\right)Y_{10}\left(\theta\right)\ ,\quad \textcolor{black}{
v_\theta^{(1)}({\bm x})=\frac{T_{10}\left(r\right)}{\sqrt{2}}\partial_\theta Y_{10}\left(\theta\right)\ ,\ {\rm and}\quad v^{(1)}_\phi({\bm x})=0}
\label{eqn:vectharm}\end{equation}
hold, where $\partial_\theta$ denotes $\partial/(\partial \theta)$ and the time variable is dropped \textcolor{black}{
because only the particular time is considered. 
Noting that the incompressibility condition gives
\begin{equation}
T_{10}=\frac{1}{r\sqrt{2}}\partial_r r^2R_{10}
\ ,\label{eqn:T10R10}\end{equation}
we eliminate \textcolor{black}{$p_{10}$} from the $r$ and $\theta$-components of
 the Stokes equation  to have
\begin{equation}
\left(\rho\partial_\rho+1\right)\left(\rho\partial_\rho-2\right)
\left(\rho \partial_\rho+3\right)\left(\rho\partial_\rho\right) {\cal R}(\rho)
=0\ ,\label{eqn:onecompcalR}\end{equation}}
where we use a dimensionless radial distance $\rho=r/a$ and
dimensionless \textcolor{black}{function,} 
\begin{equation}
    {\cal R}(\rho)= \frac{R_{10}(r)}{U} \sqrt{\frac{3}{4\pi}}
\textcolor{black}{\ .\label{eqn:defcalR}}\end{equation}

\textcolor{black}{The boundary conditions are}
\begin{equation}
  {\bm v}=\varepsilon U {\bm e}_z \ \textcolor{black}{{\rm at}}\ r=a\quad {\rm and}\quad
{\bm v}\to 0
\ {\rm as}\ r\to \infty\ .\label{eqn:bcforv}\end{equation}
\textcolor{black}{
The former represents the no-slip boundary condition at the surface, which gives $R_{10}=T_{10}/\sqrt{2}=\sqrt{4\pi/3}\  \textcolor{black}{U}$ at $r=a$.
Thus, we use (\ref{eqn:T10R10}) to obtain}
\begin{equation}
{\cal R}=1\ {\rm and}\ \partial_\rho {\cal R}=0 \ \textcolor{black}{{\rm at}} \ \rho=1
\quad {\rm and}\quad {\cal R}\to 0\ {\rm as}\ \rho\to \infty 
\ .\label{eqn:Rbc}\end{equation}
\textcolor{black}{Noting that the four operators inside the respective pairs of parentheses
in (\ref{eqn:onecompcalR}) are commutable, we find that ${\cal R}$
is given by a linear combination of $\rho^{-1}$ and $\rho^{-3}$.
The respective coefficients} \textcolor{black}{are} \textcolor{black}{found to be $3/2$ and $-1/2$ \textcolor{black}{with the aid of the surface boundary conditions in (\ref{eqn:Rbc}).}}
\textcolor{black}{We thus have} \textcolor{black}{
${\cal R}(\rho)=1+\alpha_0(\rho)$, where $\alpha_0$ is defined as}
 \begin{equation}
    \alpha_0(\rho)= \frac{3}{2\rho}-\frac{1}{2\rho^3}-1\label{eqn:defalp}
\end{equation}
\textcolor{black}{for later convenience.}

\textcolor{black}{We write $\boldsymbol{E}$ for the rate-of-strain tensor; $\textcolor{black}{E}_{xz}=(\partial_xv_z+\partial_zv_x)/2$, for example.
The drag force is along ${\bm e}_z$; its $z$ component 
is given by the surface integral of 
\begin{equation}
    {\bm e}_z\cdot\left(-p{\bm 1}+2\eta_{\rm o} \boldsymbol{E} \right)\cdot {\bm e}_r 
\label{eqn:onecompdrag}\end{equation}
over the particle surface. Here,
${\bm 1}$ denotes the isotropic tensor and ${\bm e}_r$ is the unit vector in the radial direction. 
Through the $\theta$-component of the Stokes equation at the order of $\varepsilon$,} 
\textcolor{black}{$p_{10}$} \textcolor{black}{is linked with ${\cal R}$ with the aid of (\ref{eqn:vectharm}) and (\ref{eqn:T10R10}).
We thus find the $z$-component of the drag force
at the order of $\varepsilon$ to be given by }
\begin{equation}
\textcolor{black}{-\frac{4\pi}{3}\eta_{\rm o}a\varepsilon U\left(\frac{1}{2}\partial_\rho^3+2\partial_\rho^2\textcolor{black}{-2\partial_\rho}\right){\cal R}}
\label{eqn:onecompdragz}\end{equation}
\textcolor{black}{evaluated at $\rho=1$ \citep{Fujitani2007,Okamoto-Fujitani-Komura}. 
\textcolor{black}{The term $-2\partial_\rho{\cal R}$ vanishes because of
(\ref{eqn:Rbc}).}  Substituting ${\cal R}(\rho)=1+\alpha_0(\rho)$ into the above 
and dividing the result by $\varepsilon U$ give the Stokes law.}

\textcolor{black}{
To consider how the Stokes law is modified by the preferential}  \textcolor{black}{adsorption} in the near criticality, \textcolor{black}{we
should add a term, representing the \textcolor{black}{force} caused by the composition gradient, to the Stokes equation.} 
\textcolor{black}{As shown in Sect.~\ref{sec:dyn},} \textcolor{black}{this \textcolor{black}{force}
can be calculated from the free-energy density for the mixture, which is} \textcolor{black}{
given by the integrand of the first integral of (\ref{eqn:largeF}) in the next section.}
\section{Formulation \textcolor{black}{for a near-critical binary fluid mixture}\label{sec:form}}
Suppose a near-critical binary fluid mixture with the critical composition in the homogeneous phase; $T$ is
assumed to be homogeneous and close to the critical temperature $T_{\rm c}$. 
The reduced temperature is defined as $\tau\equiv \left|T-T_{\rm c}\right|/T_{\rm c}>0$.
The composition is here represented by the difference between the mass densities of the two components, 
which is denoted by $\varphi$.  We define $\psi$ as $\varphi-\varphi_{\rm c}$, where $\varphi_{\rm c}$
represents the critical composition. 
In the mixture bulk at the equilibrium, the order parameter
$\psi$ fluctuates around zero
\textcolor{black}{significantly} on length scales smaller than the correlation length,
and the thermal average of ${\varphi}$ equals $\varphi_{\rm c}$.
On larger length scales, we can neglect the fluctuation about the \textcolor{black}{coarse-grained} equilibrium profile
\textcolor{black}{of the order parameter}, which can deviate
from zero near a wall or a surface in contact with the mixture because of the preferential  \textcolor{black}{adsorption}.
We assume the binary fluid mixture to be incompressible,
which means that the sum of the mass densities of the two components
is regarded as a \textcolor{black}{constant}.
\subsection{Statics\label{sec:statics}}
\textcolor{black}{We here assume that}
a single rigid spherical particle \textcolor{black}{is fixed in the mixture at the equilibrium
to introduce the  
$\psi$-dependent part of the free-energy functional.  It
is given by} 
\begin{equation}
F[\psi]=\int_{V^{{\rm e}}}d{\bm x}\ \left[f\left(\psi\right)+
\frac{1}{2}M(\psi) |\textcolor{black}{\boldsymbol{\textcolor{black}{\boldsymbol{\nabla}}}}\psi|^{2}\right]
+\int_{\partial V} dS\ f_{\rm s}\left(\psi\right)
\label{eqn:largeF}\ ,\end{equation}
where $\psi$ depends on the spatial position ${\bm x}$.
The first term on the right-hand side above is the volume integral over the mixture region $V^{{\rm e}}$, 
while the second term is the surface integral
over the particle surface $\partial V$.   \textcolor{black}{The definitions of $f(\psi)$ and $M(\psi)$ are given below.} 
The surface energy density, denoted by $f_{\rm s}(\psi)$, is assumed to be linear with respect to $\psi$; 
we suppose that the preferential  \textcolor{black}{adsorption} is caused by 
such a short-range interaction as the hydrogen bond. The surface field is defined as ${h}=
-f'_{\rm s}(\psi)$; the prime hereafter indicates the derivative with respect to the variable.
In the Gaussian model, 
$f(\psi)$ is a quadratic function and $M$ is a constant. In the present study, \textcolor{black}{as mentioned in Sect.~\ref{sec:intro},
we employ 
the renormalized local functional theory \citep{OkamotoOnuki}, where $f$ is given by}
\begin{eqnarray}
&& f(\psi)\equiv k_{\rm B}T_{\rm c} \left( \frac{1}{2} C_1\xi_0^{-2}\omega^{\gamma-1}\tau\psi^2 +
\frac{1}{12} C_1C_2\xi_0^{-2} \omega^{\gamma-2\beta} \psi^4 \right)\ .
\label{eqn:f}\end{eqnarray}
\textcolor{black}{This is obtained after coarse-graining up to the local correlation length, for which we write $\xi$}.
Hereafter, $\alpha,\beta,\gamma,\nu,$ and $\eta$ are the critical exponents 
for binary mixtures near the demixing critical point (or in the three-dimensional Ising model), $C_1$ is a \textcolor{black}{positive} 
nonuniversal constant,  and $C_{2}$ is given by
\begin{equation}
C_{2}=3u^{*}C_{1}\xi_{0}.
\label{eqn:defC2}\end{equation}
Here, \textcolor{black}{$\xi_0$ denotes a nonuniversal microscopic length, and}
\textcolor{black}{$u^{*}$ denotes the scaled coupling constant  at the Wilson-Fisher fixed point. We have
$u^{*}=2\pi^{2}/9$ in the three dimensions at the $1$-loop order.}
The "distance" from the criticality is represented by a dimensionless quantity $\omega$, which is defined to give
\begin{equation}
\textcolor{black}{\xi=\xi_0 \omega^{-\nu}}\ .\label{eqn:xiomega}\end{equation}
A self-consistent condition gives
\begin{equation}
\omega=\tau+C_{2}\omega^{1-2\beta}\psi^{2}\ ,
\label{eqn:omega}\end{equation}
which leads to $\xi_\infty=\xi_0\tau^{-\nu}$ 
because the composition is critical far from the particle. 
The coefficient of the square gradient term in (\ref{eqn:largeF}) equals 
\begin{equation}
M=k_{\rm B}T_{\rm c} C_{1}\omega^{-\eta\nu}
\ .\label{eqn:Mdef}\end{equation}
\textcolor{black}{Thus, $M$ is a positive function of $\psi$ at a given temperature. }

\textcolor{black}{The order-parameter fluctuation is significant only on the length scales smaller than $\xi$.
On larger \textcolor{black}{length-}scales, the probability distribution of the order-parameter profile should have
a sharp peak around its maximum, and thus  
the most probable profile is regarded as observed without fluctuation 
\citep{OkamotoOnuki, Yabu-Oka-On, Yabu-On}.}  \textcolor{black}{Hence, in the renormalized local functional theory,
we can obtain this profile by minimizing} (\ref{eqn:largeF}),
which is regarded as the grand-potential functional with the chemical potential $\mu$,
conjugate to $\psi$, being put equal to zero; \textcolor{black}{$\mu$ vanishes} because of the critical composition
$(\psi=0$) far from the particle.
We thus find the equilibrium profile to satisfy 
\begin{equation}
    0=f'(\psi)-\frac{1}{2}M'(\psi)|\textcolor{black}{\boldsymbol{\textcolor{black}{\boldsymbol{\nabla}}}} \psi|^2-M(\psi)\Delta\psi
\quad  {\rm in} \ V^{\rm e} \ ,\label{eqn:mu0}\end{equation}  \textcolor{black}{whose left-hand side implies
$\mu=0$, and }
\begin{equation}
0={h}+ M{\bm n}\cdot \textcolor{black}{\boldsymbol{\textcolor{black}{\boldsymbol{\nabla}}}}\psi \quad  {\rm at} \ \partial V\ ,
\label{eqn:psiatsurface}\end{equation}
where ${\bm n}$ is the unit normal vector on the surface towards outside the particle. 
\textcolor{black}{The equilibrium profile can deviate from zero near the particle surface because of the preferential  \textcolor{black}{adsorption}.}

We here assume that the preferential  \textcolor{black}{adsorption} is represented by only the surface field
and neglect higher-order terms with respect to $\psi$, 
such as the second-order term involving the surface enhancement \citep{bray, diehl86, diehl97, cardy},
to study how the adsorption influences the drag coefficient, as in the previous studies 
of the renormalized local functional theory or the deviation
of the drag coefficient \citep{OkamotoOnuki, Yabu-Oka-On, Yabu-On, Yabu-Oka-On-Soft, Okamoto-Fujitani-Komura, furu,  Fujitani2018}.  
\subsection{Dynamics\label{sec:dyn}}
\textcolor{black}{
In the equilibrium, like an inhomogeneous magnetic field for a magnetic system, $\mu$ can be assumed to depend on ${\bm x}$. Under inhomogeneous chemical potential, assuming no bulk phase separation, still we can obtain the coarse-grained equilibrium profile by minimizing the grand-potential functional.}
\textcolor{black}{
Conversely, for a
given coarse-grained profile $\psi({\bm x})$, (\ref{eqn:largeF}) gives the free-energy functional, and
$\mu({\bm x})$ is given by its
functional derivative  with respect to $\psi({\bm x})$}. 

In the dynamics \textcolor{black}{with} the local equilibrium, 
the chemical potential $\mu$ is \textcolor{black}{thus} still given by the left-hand side of (\ref{eqn:mu0}), {\it i.e.\/},
\begin{equation}
    \mu({\bm x},t)=f'(\psi)-\frac{1}{2}M'(\psi)|\textcolor{black}{\boldsymbol{\textcolor{black}{\boldsymbol{\nabla}}}} \psi|^2-M(\psi)\Delta\psi
\quad  {\rm for} \ {\bm x}\in V^{\rm e} \ ,\label{eqn:mu}\end{equation}
and (\ref{eqn:psiatsurface}) holds as it is. 
The latter is mentioned
at (62c) of \citep{DJ} and in Appendix A of \citep{Okamoto-Fujitani-Komura}.
Hereafter, $\psi({\bm x},t)$ represents the coarse-grained 
profile of the order parameter.

\textcolor{black}{Considering the change in the free-energy functional
due to the quasi-static deformation of the mixture, we can obtain the reversible part of the
stress tensor \citep{Onuki-1}, as is done in the model H.  }
\textcolor{black}{Writing $\bm{\Pi}$} for its negative, we have
\begin{equation}
\textcolor{black}{\bm{\Pi}}=\left(-f+\mu\psi-\frac{M}{2}\left|\textcolor{black}{\boldsymbol{\textcolor{black}{\boldsymbol{\nabla}}}}\psi\right|^2 \right){\bm 1}+M\textcolor{black}{\boldsymbol{\textcolor{black}{\boldsymbol{\nabla}}}}\psi\textcolor{black}{\boldsymbol{\textcolor{black}{\boldsymbol{\nabla}}}}\psi
\ .\label{eqn:pidef}\end{equation}
\textcolor{black}{Without the composition gradient, the above} gives a better-known
expression of the osmotic pressure exerted on a semipermeable membrane $\psi f'-f$, which is mentioned
in \citep{deg} for example.
The velocity field in the mixture ${\bm v}$ 
still satisfies the first entry of (\ref{eqn:onecomp}). 
Because of
\begin{equation} 
\textcolor{black}{\boldsymbol{\textcolor{black}{\boldsymbol{\nabla}}}}\cdot \textcolor{black}{\bm{\Pi}}=\psi \textcolor{black}{\boldsymbol{\textcolor{black}{\boldsymbol{\nabla}}}} \mu
\label{eqn:nabpsi}\ ,\end{equation} 
we can regard ${\bm v}$ as satisfying
\begin{equation}
0=-\textcolor{black}{\boldsymbol{\textcolor{black}{\boldsymbol{\nabla}}}} {p}-\psi \textcolor{black}{\boldsymbol{\textcolor{black}{\boldsymbol{\nabla}}}}\mu +\eta_{\rm o} \Delta {\bm v}\ ,
\label{modelH2}\end{equation}
instead of the second entry of (\ref{eqn:onecomp}).
\textcolor{black}{The inertia term vanishes for the same reason as mentioned
in Sect.~\ref{sec:prel};}
the pressure ${p}$ \textcolor{black}{still} plays a role of keeping the incompressibility. 
The viscosity $\eta_{\rm o}$ is assumed to be homogeneous,
considering the weak dependence of its singular part
on $\xi$ \citep{ohta, ohtakawasaki}.  \textcolor{black}{This point is discussed in Sect.~\ref{sec:discuss}.}
The diffusive flux between the two components is proportional to the gradient of $\mu$, 
and the mass conservation of each component leads to   
\begin{equation}
\frac{\partial}{\partial t}\psi ({\bm x}, t)
=-{\bm v}\cdot \textcolor{black}{\boldsymbol{\textcolor{black}{\boldsymbol{\nabla}}}}\psi+ \textcolor{black}{\boldsymbol{\textcolor{black}{\boldsymbol{\nabla}}}} \cdot\left[L(\psi) \textcolor{black}{\boldsymbol{\textcolor{black}{\boldsymbol{\nabla}}}} \mu\right].
\label{modelH1}\end{equation}
The first and second terms on the right-hand side above represent the mass transport
due to the convection and the one due to the diffusion, respectively.  
The Onsager coefficient $L(>0)$, \textcolor{black}{depending} on $\xi$, \textcolor{black}{is} regarded as a function of $\psi$ at a given temperature,
\textcolor{black}{and is further
discussed in Sect.~\ref{sec:dep}}.
\textcolor{black}{Equations (\ref{modelH2}) and (\ref{modelH1}), \textcolor{black}{together with the first entry of (\ref{eqn:onecomp}}), describe
the hydrodynamics of the mixture on length scales larger than \textcolor{black}{$\xi$}
\citep{Yabu-Oka-On-Soft,undul}}.

\textcolor{black}{As in Sect.~\ref{sec:prel},} we assume that the particle
undergoes translational motion with a constant velocity $\varepsilon U{\bm e}_z$ in a quiescent fluid,
\textcolor{black}{and proceed with calculations without making the physical meaning of $\varepsilon$ explicit.
A possible meaning is discussed in Sect.~\ref{sec:discuss}.  
The boundary conditions for ${\bm v}$ are the same as described in Sect.~\ref{sec:prel}.
The diffusion flux} cannot pass across the particle surface, which leads to
\begin{equation}
L\partial_r \mu=0  \ {\rm for}\ r=a\quad {\rm and}\quad \mu \to 0 \ {\rm as}\ r\to \infty\ .
\label{eqn:bcformu}\end{equation}
The first entry of (\ref{eqn:bcformu}) comes because the diffusive flux is defined on the 
frame co-moving with the particle.
The surface boundary condition for $\psi$ is given by (\ref{eqn:psiatsurface}); $\psi$ approaches zero far from the particle.
This and the second entry of (\ref{eqn:bcformu}) come because of the critical composition far from the particle. 
Viewed from the co-moving frame, the fields are stationary, and thus (\ref{modelH1})
leads to
\begin{equation}
\left[{\bm v}-\varepsilon U {\bm e}_z\right] \cdot \textcolor{black}{\boldsymbol{\textcolor{black}{\boldsymbol{\nabla}}}}\psi= \textcolor{black}{\boldsymbol{\nabla}} \cdot \left[L(\psi) \textcolor{black}{\boldsymbol{\nabla}} \mu\right].
\label{modelH1st}\end{equation}
\subsection{Set up for calculations \label{sec:perturb}}
\textcolor{black}{In the reference state at $\epsilon=0$,}  the particle center is fixed at the origin and the ambient mixture is quiescent, \textcolor{black}{
as in \citep{Fujitani2018}.}  In this state, $\psi$ can be calculated from (\ref{eqn:mu0}) and (\ref{eqn:psiatsurface}) with $\partial V$
being located at $r=a$; we write $\psi^{(0)}(r)$ for this solution, considering  the \textcolor{black}{rotational} symmetry of the
reference state. Over the mixture in this state, $\mu$ vanishes
and $p$ equals a constant, for which we write ${p}^{(0)}$.  \textcolor{black}{In addition to (\ref{eqn:vpexp}), }\textcolor{black}{ we define $\psi^{(1)}$ and $\mu^{(1)}$ so that} 
\begin{equation}
\psi({\bm x})=\psi^{(0)}(r)+\varepsilon 
\psi^{(1)}({\bm x}) \textcolor{black}{\quad{\rm and}
\quad \mu({\bm x})=\varepsilon \mu^{(1)}({\bm x})\label{eqn:perexp}}\end{equation}
\textcolor{black}{hold up to the order of $\varepsilon$. The fields with the superscript $^{(1)}$
all vanish far from the particle.} 
The time variable $t$ is not explicitly written here \textcolor{black}{for the same reason as mentioned in Sect.~\ref{sec:prel}}.
Because of the symmetry of the particle motion, we have \textcolor{black}{(\ref{eqn:defp10}),
 (\ref{eqn:vectharm}),} and
\begin{equation}
\mu^{\left(1\right)}\left({\bm x} \right)=Q_{10}\left(r\right)Y_{10}\left(\theta\right)\ 
,\label{lin-chemp-radial}\end{equation}
whereby $Q_{10}$ \textcolor{black}{is} defined.  
\textcolor{black}{In addition to (\ref{eqn:defcalR}),} we introduce
dimensionless functions \begin{equation}
    {\cal Q}(\rho)= \frac{Q_{10}(r)\sqrt{L(0)}}{U \sqrt{\eta_{\rm o}}}  \sqrt{\frac{3}{20\pi}}\ ,\label{eqn:defRQ}
    \end{equation}
and \begin{equation}
    \Psi(\rho)=- \frac{r^2}{3\sqrt{5\eta_{\rm o} L(0)}} \frac{d\psi^{(0)}(r)}{dr}\ ,\label{eqn:defPsi}
\end{equation}
where $L(0)$ is $L(\psi)$ for $\psi=0$.
We rewrite
the $r$ and $\theta$ components of (\ref{modelH2}) at the order of $\varepsilon$ by using 
(\ref{eqn:vpexp})\textcolor{black}{--}(\ref{eqn:vectharm}), (\ref{eqn:perexp}) and (\ref{lin-chemp-radial}).
\textcolor{black}{Using the same procedure that leads to (\ref{eqn:onecompcalR}), we here instead obtain}
\begin{equation}
\left(\rho\partial_\rho+1\right)\left(\rho\partial_\rho-2\right)
\left(\rho \partial_\rho+3\right)\rho\partial_\rho {\cal R}(\rho)
=-30 \Psi(\rho) {\cal Q}(\rho)\ .
\label{eqn:R10dless}\end{equation}
Equation (\ref{modelH1st}) yields
\begin{equation}
\left( \rho\partial_\rho-1\right)\left(\rho\partial_\rho+2\right) {\cal Q}(\rho)
=-3\Psi(\rho)\left[A(\rho)\left({\cal R}(\rho)-1\right) -B(\rho)\partial_\rho {\cal Q}(\rho)
 \right]
\ ,\label{eqn:Q10dless}\end{equation}
where $A$ and $B$ are defined as
\begin{equation} A(\rho)=\frac{L(0)}{L(\psi^{(0)}(a\rho))}\quad {\rm and}\quad B(\rho)= \frac{L'(\psi^{(0)}(a\rho))}{a L(\psi^{(0)}(a\rho))}
\sqrt{5\eta_{\rm o}L(0)}\ .
\end{equation}
\textcolor{black}{We have (\ref{eqn:Rbc})},
while (\ref{eqn:bcformu}) gives
\begin{equation}
   \partial_\rho {\cal Q}=0 \ {\rm for} \ \rho=1
\quad {\rm and}\quad {\cal Q}\to 0\ {\rm as}\ \rho\to \infty  \ .\label{eqn:Qbc}\end{equation}
Regarding (\ref{eqn:R10dless}) as an equation for ${\cal R}$, we use \textcolor{black}{
(\ref{eqn:Rbc}) to }have
\begin{equation}
 {\cal R}\left(\rho\right)=
1+\alpha_0(\rho)+\int_{1}^{\infty}d\sigma\ \varGamma_{R}\left(\rho,\sigma\right)\Psi\left(\sigma\right)
{\cal Q}\left(\sigma\right)\ ,
\label{inteq-R}\end{equation}
where the kernel $\varGamma_{R}$ is given in Appendix
and $\alpha_0$ is defined \textcolor{black}{by (\ref{eqn:defalp}).
Without the preferential  \textcolor{black}{adsorption}, $\Psi$ and ${\cal Q}$ vanishes, and
${\cal R}$ becomes $1+\alpha_0$, as described in Sect.~\ref{sec:prel}. 
Regarding (\ref{eqn:Q10dless}) as an equation for ${\cal Q}$, we use
(\ref{eqn:Qbc}) to have
\begin{equation}
{\cal Q}\left(\rho\right)=\int_{1}^{\infty}d\sigma\ \varGamma_{Q}\left(\rho,\sigma\right)\Psi(\sigma)\left[
A(\sigma)\left({\cal R}(\sigma)-1\right) -B(\sigma)\partial_\sigma {\cal Q}(\sigma)\right]
\ ,\label{inteq-Q}
\end{equation}
where the kernel  $\varGamma_{Q}$ is shown in Appendix.}

\textcolor{black}{\textcolor{black}{Instead of  the integral of (\ref{eqn:onecompdrag}) over the particle surface},
the $z$ component of the drag force
is given by} \citep{effvis, rigid}
\begin{equation}
    \int_{\partial V}dS\ {\bm e}_z\cdot\left(- \textcolor{black}{\textcolor{black}{\boldsymbol{\Pi}}}\cdot{\bm e}_r
  +\textcolor{black}{\boldsymbol{\nabla}}_\parallel f_{\rm s}-\frac{2f_{\rm s}}{a}{\bm e}_r -p{\bm 1} \textcolor{black}{\cdot{\bm e}_r}+2\eta_{\rm o} \boldsymbol{E} \cdot {\bm e}_r\right)\ ,
\label{eqn:dragintegral}\end{equation}
where 
$\textcolor{black}{\boldsymbol{\nabla}}_\parallel$ denotes the projection of $\textcolor{black}{\boldsymbol{\nabla}}$ on the tangent plane and $1/a$ is the mean curvature of the particle surface.  
The two terms involving $f_s$ 
in (\ref{eqn:dragintegral}) are canceled with each other
after the integration, and thus need not be considered here,
as mentioned in Appendix A of \citep{effvis}.
Writing $\varepsilon  \textcolor{black}{\textcolor{black}{\boldsymbol{\Pi}}}^{(1)}$ for $\textcolor{black}{\boldsymbol{\Pi}}$ at the order of ${\varepsilon}$, we need to calculate
the left-hand side of (\ref{eqn:piforce}) below in
calculating the contribution from the first term in the parentheses of (\ref{eqn:dragintegral}) to $\gamma_{\rm d}$. 
Because of (\ref{eqn:pidef}), $\textcolor{black}{\boldsymbol{\Pi}}^{(1)}$ contains
$\psi^{(1)}$, which satisfies \textcolor{black}{(\ref{eqn:psiatsurface}) and }(\ref{eqn:mu}).  \textcolor{black}{It is natural that}
the distortion of the order-parameter profile from the equilibrium one
\textcolor{black}{should} contribute to $\Delta\gamma_{\rm d}$. Thus, it appears that, even after we have solved
$\mathcal{R}(\rho)$ and $\mathcal{Q}(\rho)$ for a given $\Psi(\rho)$ to find \textcolor{black}{${\bm v}^{(1)}$ and $\mu^{(1)}$}, 
we still have to obtain $\psi^{(1)}$
from \textcolor{black}{(\ref{eqn:psiatsurface}) and} (\ref{eqn:mu}) in advance to calculate  $\gamma_{\rm d}$.  
\textcolor{black}{However, this calculation can be evaded.} We have
\begin{eqnarray}&&
    \int_{\partial V} dS\ {\bm e}_z\cdot \left(-\textcolor{black}{\boldsymbol{\Pi}}^{(1)}\cdot{\bm e}_r\right)=
     \int_{V^{{\rm e}}} d{\bm x}\ \psi^{(0)}\partial_z \mu^{(1)} \nonumber\\&&\qquad=
     -\int_{\partial V} dS\ \left[\mu^{(1)}\psi^{(0)}\right] \cos{\theta}
    -\int_{V^{{\rm e}}} d{\bm x}\ \mu^{(1)}{\psi^{(0)}}' \cos{\theta}\ ,\label{eqn:piforce}
\end{eqnarray}
where the two equalities are derived with the aid of
(\ref{eqn:nabpsi}) and Gauss' divergence theorem. 
The formula given by (\ref{eqn:piforce}) is helpful
because its right-hand side does not contain $\psi^{(1)}$ and  
we need not obtain $\psi^{(1)}$ from \textcolor{black}{(\ref{eqn:psiatsurface}) and} (\ref{eqn:mu}).
The formula is essentially the same as used at (37) of \citep{Fujitani2018}, 
where the author calculated the quantity corresponding with $\psi^{(1)}$ by solving a linear differential equation to derive the formula
in the Gaussian model.  This derivation cannot be applied in the present formulation because
an analytical expression of $\psi^{(1)}$ cannot be derived from (\ref{eqn:mu}), which is highly nonlinear
with respect to $\psi$.
Equation (\ref{eqn:piforce}) gives an alternative and much simpler derivation, which is valid
in cases more general than 
the previous derivation, including the present formulation.

The rest terms in the integrand of  (\ref{eqn:dragintegral})  involve $p$ and $\boldsymbol{E}$. 
The latter is rewritten in
terms of ${\cal R}$ through the definition of $\boldsymbol{E}$. 
The former at the order of $\varepsilon$
is related to ${\cal R}$ with the aid of the $\theta$
component of (\ref{modelH2}), \textcolor{black}{and contains}  \textcolor{black}{the term coming from $ -\psi \textcolor{black}{\boldsymbol{\nabla}}\mu$,
\textcolor{black}{which} is canceled by the first term on the right-hand side of (\ref{eqn:piforce}) in calculating (\ref{eqn:dragintegral})
at the order of $\varepsilon$.}  \textcolor{black}{Thus we can use  (\ref{inteq-R}) and (\ref{eqn:dragintegral}) to obtain }
\begin{equation}
\gamma_{\rm d}=6\pi\eta_{\rm o}a\left( 1+ \frac{10}{3} \int_1^\infty d\rho\ 
\alpha_0(\rho) {\cal Q}(\rho) \Psi(\rho)\right)
\label{eqn:draco}\end{equation}
\textcolor{black}{with the aid of (\ref{eqn:alas}).} The second term in the parentheses above gives $\Delta\gamma_{\rm d}$ divided by $6\pi\eta_{\rm o}a$ of the Stokes law.  The quotient is below referred to as the 
normalized deviation, denoted by $\Delta{\hat \gamma}_{\rm d}$.  Instead of using (\ref{eqn:alas}),
we can use the Lorentz reciprocal theorem
\citep{lorentz} to derive (\ref{eqn:draco}), as shown in Appendix B of \citep{Fujitani2018}.  
\section{Elements of the calculation procedure\label{sec:calc}}
Our task is to calculate the drag coefficient by means of (\ref{eqn:draco}).  For this, we should calculate ${\cal Q}$ by 
determining the dependence of $L$ on $\psi$ and by obtaining $\psi^{(0)}$.  
How to carry out these steps are mentioned in the following subsections.
\textcolor{black}{To obtain ${\cal Q}$, we should solve the simultaneous equations with respect to ${\cal R}$ and ${\cal Q}$,
(\ref{inteq-R}) and (\ref{inteq-Q}).
Simultaneous} equations of the same type are derived in the Gaussian model in \citep{Fujitani2018},
where the derivative of the difference in the mass densities
in the reference state, corresponding with ${\psi^{(0)}}'$ in the present study, 
is proportional to the surface field and a series expansion with respect to a dimensionless surface field is naturally introduced.
In contrast, in the present formulation,
$\Psi\propto {\psi^{(0)}}'$ in (\ref{inteq-R}) and (\ref{inteq-Q}) is not proportional to ${h}$.
Thus, in Sect.~\ref{sec:calQ}, we introduce an artificial parameter to
derive a series expansion of the solution.  

\subsection{Calculation for ${\cal Q}$ \label{sec:calQ}}
\textcolor{black}{We modify (\ref{inteq-R}) and (\ref{inteq-Q}) by replacing $\Psi$ with $\kappa \Psi$, where $\kappa$ is the
artificial parameter.  The solutions of these modified equations
become dependent on $\kappa$, and are denoted by  ${\tilde {\cal R}}(\rho, \kappa)$ and ${\tilde {\cal Q}}(\rho, \kappa)$.  
The original solutions, ${\cal R}(\rho)$ and ${\cal Q}(\rho)$, are respectively equal to  ${\tilde {\cal R}}(\rho, 1)$ and ${\tilde {\cal Q}}(\rho, 1)$.
Substituting the modified equation from (\ref{inteq-R}), {\it i.e.\/},
\begin{equation}
{\tilde {\cal R}}\left(\rho,\kappa\right)=
1+\alpha_0(\rho)+\kappa \int_{1}^{\infty}d\sigma\ \varGamma_{R}\left(\rho,\sigma\right)\Psi\left(\sigma\right)
{\tilde {\cal Q}}\left(\sigma,\kappa\right)\ ,
\label{inteq-R1}\end{equation}
into the modified equation from (\ref{inteq-Q}), {\it i.e.\/},}
\begin{equation}
{\tilde {\cal Q}}\left(\rho, \kappa\right)=\kappa\int_{1}^{\infty}d\sigma\ \varGamma_{Q}\left(\rho,\sigma\right)\Psi(\sigma)\left[
A(\sigma)\left({\tilde {\cal R}}(\sigma,\kappa)-1\right) -B(\sigma)\partial_\sigma {\tilde {\cal Q}}(\sigma,\kappa)\right]
\ ,\label{inteq-Q1}
\end{equation}
\textcolor{black}{gives an integral equation with respect to ${\tilde Q}$.}

Writing ${\tilde q}_1, {\tilde q}_2, \ldots$ for the expansion coefficients,
we assume
\begin{equation}
{\tilde{\cal Q}}\left(\rho, \kappa\right)=\sum_{n=1}^{\infty} {\tilde q}_{n}\left(\rho \right)\kappa^{n}
\textcolor{black}{\ ,\label{seriesQmod}}
\end{equation}
\textcolor{black}{which is substituted into the integral equation with respect to ${\tilde Q}$
to yield the recurrence relations for the expansion coefficients.}
The difference in the braces of (\ref{inteq-Q1}) 
equals $A(\sigma)\alpha_0(\rho)$ at the order of $\kappa^0$, and we obtain
\begin{equation}
    {\tilde q}_1(\rho)= \int_1^\infty d\sigma\ \varGamma_{Q}\left(\rho,\sigma\right)\Psi(\sigma)A(\sigma)  \alpha_0(\sigma)
\ .\label{eqn:q1}\end{equation}
Similarly, we \textcolor{black}{pick up the terms at the order of $\kappa^n$ $(n=2,3,\ldots)$ from the integral equation to} obtain
\begin{equation}
{\tilde q}_2\left(\rho\right)=-\int_{1}^{\infty}d\sigma\varGamma_{Q}\left(\rho,\sigma\right) 
\Psi\left(\sigma\right) B(\sigma)\frac{\partial}{\partial\sigma}{\tilde q}_1\left(\sigma\right)\ ,
\label{eqn:q2}\end{equation}
and  
\begin{equation}
{\tilde q}_{n}\left(\rho\right)=\int_{1}^{\infty}d\sigma\left[\varGamma\left(\rho,\sigma\right)
\Psi\left(\sigma\right)q_{n-2}\left(\sigma\right)-\varGamma_{Q}\left(\rho,\sigma\right)
\Psi\left(\sigma\right)B(\sigma)\frac{\partial}{\partial\sigma}q_{n-1}\left(\sigma\right)\right]
\label{eqn:qn}\end{equation}
for $n=3,4,\ldots$, 
where the kernel $\varGamma$ is defined as
\begin{equation}
\varGamma\left(\rho,\sigma\right)=\int_{1}^{\infty}d\tau\ \varGamma_{Q}\left(\rho,\tau\right)
\Psi\left(\tau\right)A(\tau)
\varGamma_{R}\left(\tau,\sigma\right).
\end{equation}
\textcolor{black}{Substituting (\ref{seriesQmod}) with $\kappa=1$ into (\ref{eqn:draco}) gives  
\begin{equation}
\Delta{\hat \gamma}_{\rm d}=\sum_{n=1}^\infty \chi_n\ ,
\label{eqn:chichi}\end{equation}} where
$\chi_n$ is defined as
\begin{equation}
\chi_n= \frac{10}{3} \int_1^\infty{\rm d}\rho\ \alpha_0\textcolor{black}{(\rho)}{\tilde q}_n(\rho)\Psi(\rho)\ .\label{eqn:chidef}
\end{equation}

 \subsection{Dependence of $L$ on the order parameter \label{sec:dep}}
In this paragraph, 
we consider the near-critical fluctuation 
about the equilibrium in the absence of a particle; $\xi$ is homogeneous.
Here, the thermal average of the composition in the mixture is uniform but not necessarily equal to the critical one;
$\varphi({\bm x},t)$ represents the fluctuating composition and
$\delta\varphi({\bm x}, t)$ is defined as  $\varphi({\bm x},t)-{\bar\varphi}$, where ${\bar\varphi}$ denotes the thermal 
average of $\varphi$.  We write
$C_{\bm k}(t)$ for the spacial Fourier transform of the correlation function, {\it i.e.\/},
the thermal average of $\delta\varphi({\bm x}, t)\delta\varphi({\bm 0}, 0)$, with ${\bm k}$ denoting the wave-number vector.
In the mode-coupling theory \citep{kawasaki, swin, Onuki-1}, the convection at smaller length scales is regarded as a part of the diffusion at larger length scales, which leads to
\begin{equation}
    \frac{\partial }{\partial t}C_{\bm k}(t)=\textcolor{black}{-}\frac{k_{\rm B} T_{\rm c}}{6\pi \eta_{\rm o} \xi} \left|{\bm k}\right|^2 C_{\bm k}(t)
\label{eqn:mct}\end{equation}
for small wave-number $|{\bm k}|\stackrel{<}{\sim} \xi^{-1}$. 
\textcolor{black}{ Here, the regular part 
 is assumed to be much smaller than the singular part.} 
The fraction in (\ref{eqn:mct}) gives  the diffusion coefficient in a coarse-grained picture, where
the convection does not contribute to the mass transport.  
The time derivative of $\delta\varphi$ on large length-scales should be also given by
the second term on the right-hand side
of (\ref{modelH1}) \textcolor{black}{with $L(\psi)$ being regarded as $L({\bar\varphi})$.}
Then, we can approximate $\mu$ of (\ref{eqn:mu}) to be $f'(\psi)$
for long wave-length fluctuations, and rewrite 
the second term as $Lf''(\psi)\Delta \delta{\varphi}({\bm x},t)$.  
Multiplying this term with ${\delta\varphi}({\bm 0},0)$ and taking the thermal average of the product, 
we should find that the Fourier transform of the average equals the right-hand side of (\ref{eqn:mct}).  
For this to hold, $Lf''$ should be approximately equal to the fraction in (\ref{eqn:mct}) for 
$|{\bm k}|\xi\ll 1$.

Let us turn back to our calculation of the drag coefficient. We assume that
the equality mentioned at the end of the preceding paragraph locally holds when the particle moves in the mixture.  
Thus, in calculating $\gamma_{\rm d}$, we use
\begin{equation}
    L(\psi)= \frac{k_{\rm B}T}{6\pi\eta_{\rm o}\xi f''(\psi)} \ ,\label{eqn:Ldef}
\end{equation}
where $f''$ is the inverse susceptibility. 
The approximation introduced here presupposes small
spatial variation of $\psi$.  At least, the variation of $L(\psi)$ over $\xi$ is required to be small as compared with its typical value
at every locus.
From (\ref{eqn:f}) and (\ref{eqn:omega}), we derive
\begin{equation}
f'(\psi)=k_{B}T_{c}C_{2}\psi\omega^{\gamma}
\frac{2-\alpha+4\left(1-\alpha\right)
\tau\omega^{-1}+5\alpha {\tau^2\omega^{-2}}}{18u^{*}\left[2\beta+\left(1-2\beta\right)
{\tau\omega^{-1}}\right]\xi_{0}^{{3}}}
\ .\label{eqn:fpp}\end{equation}
Similarly, we can calculate 
$f''(\psi)$, although its lengthy expression is not shown here. This expression tells
\begin{equation}
    f''(\psi)\to k_{B}T_{c}C_1\tau^\gamma\xi_0^{-2}
    \quad {\rm as}\ \psi\to 0\ ,
\label{eqn:suslim}\end{equation}
which leads to 
\begin{equation}
    L(0)=\frac{\xi_0}{6\pi\eta_{\rm o} C_1\tau^{\gamma-\nu}}
\ .\label{eqn:lzero}\end{equation}
The scaling relation for the critical exponents gives $\gamma\approx 2\nu$ because 
\textcolor{black}{of $\eta\approx 0.036\ll 2$ for three-dimensional binary fluid \citep{peli}}, and thus 
$L(0)$ is approximately proportional to $\xi_\infty$.
\textcolor{black}{In more general, considering that $f''$ gives the inverse of the susceptibility 
involving the critical exponent $\gamma$, the scaling relation and (\ref{eqn:Ldef}) approximately lead to
$L\propto \xi$. } \textcolor{black}{This linearity is expected in view of
the mode-coupling theory, although the power} \textcolor{black}{is found to be slightly smaller than unity 
in the dynamic renormalization group calculation \citep{sigg}.}

\subsection{{Dimensionless parameters and
the} equilibrium profile \label{sec:nond}}
We numerically calculate the equilibrium profile, $\psi^{(0)}(r)$, by solving
(\ref{eqn:mu0}) and (\ref{eqn:psiatsurface}) \textcolor{black}{in combination with (\ref{eqn:omega})}, as was done in \citep{OkaOnPRE}.
\textcolor{black}{Equation (\ref{eqn:xiomega}) then gives $\xi$.}
\textcolor{black}{Some dimensionless parameters are introduced below to facilitate numerical calculations.} 
A characteristic reduced temperature $\tau_{\rm a}$ is defined so that $\xi$ becomes $a$ for $\psi=0$ at $\tau=\tau_a$.  
A characteristic order parameter $\psi_{\rm a}$ is defined so that
$\xi$ becomes  $a$  for $\psi=\psi_{\rm a}$ 
at $\tau=0$.  Equation (\ref{eqn:omega}) gives
\begin{equation}
\tau_{\rm a}=\left(\xi_{0}/a\right)^{1/\nu}\quad {\rm and}\quad 
\psi_{\rm a}=\tau_{\rm a}^\beta/\sqrt{C_2}
\ .\label{eqn:nodim}\end{equation}
The reduced temperature and order parameter are respectively scaled as
${\hat \tau}\equiv \tau/\tau_a$ and ${\hat \psi}(\rho)\equiv \psi(a\rho )/\psi_{\rm a}$.
Introducing a dimensionless surface field, defined as
\begin{equation}
    {\hat h}= \frac{{h}a\sqrt{C_2}}{k_{\rm B}T_{\rm c}C_1\tau_{\rm a}^\beta} 
\ ,\label{eqn:hath}\end{equation}
we can rewrite (\ref{eqn:psiatsurface}) as
\begin{equation}
    \partial_\rho \hat{\psi} ^{(0)} (\rho)=-{\hat h} \omega^{\nu\eta}\quad {\rm at}\ \rho=1
\ ,\label{eqn:boundcond}\end{equation}
where ${\hat\psi}^{(0)}(\rho)$ is defined as $\psi^{(0)}(a\rho )/\psi_{\rm a}$. Noting   \textcolor{black}{(\ref{eqn:defPsi})} and
(\ref{eqn:lzero}), we find 
\begin{equation}
    \Psi(\rho)=-\frac{\rho^2}{\sqrt{5\pi {\hat \tau}^{\nu-\gamma}}} \frac{d {\hat \psi}^{(0)}(\rho)}{d\rho}
\ .\label{eqn:Psipsi}\end{equation}

We \textcolor{black}{use $\nu=0.627$; see \citep{peli} for the values of the critical exponents.}  
Unless otherwise stated, we hereafter
put $\eta=0$ for simplicity, \textcolor{black}{considering its small positive value mentioned above}.  This allows us to take 
$M$ as a constant $k_{\rm B}T_{\rm c}C_1$ because of (\ref{eqn:Mdef}).  
Equations (\ref{eqn:mu0}) and (\ref{eqn:fpp}) give
\begin{eqnarray}&&
    \left( \partial_\rho^2+ 2\rho^{-1}\partial_\rho \right){\hat \psi}^{(0)}(\rho)\nonumber\\
    &&\quad ={
    \frac{\left[2-\alpha+4\left(1-\alpha\right){\hat\tau}{\hat\omega}^{-1}+5\alpha {\hat\tau}^2{\hat\omega}^{-2}\right]{\hat \omega}^\gamma}{6\left[2\beta+\left(1-2\beta\right){\hat\tau}{\hat\omega}^{-1}\right]}   {\hat \psi}^{(0)}(\rho) }
    \quad {\rm for}\ \rho>1 \ ,
\label{eqn:eqforprofile}\end{eqnarray}
where ${\hat \omega}$ is defined as
$\omega/\tau_{\rm a}$ and satisfies
${\hat \omega}={\hat\tau}+{\hat\omega}^{1-2\beta}{\hat\psi}^2$ because of (\ref{eqn:omega}).
We numerically solve (\ref{eqn:boundcond})
with $\eta$ being put equal to zero and (\ref{eqn:eqforprofile}) by using Mathematica (Wolfram)
to obtain ${\hat \psi}^{(0)}$, and then
$\Psi(\rho)$ with the aid of (\ref{eqn:Psipsi}).
When $\eta$ vanishes, (\ref{eqn:boundcond}) does not involve $\omega$, and thus
we can proceed with the calculation only by fixing the values of ${\hat h}$ and $\xi_\infty/a={\hat \tau}^{-\nu}$. 
Otherwise, the value of $\tau_{\rm a}$ should be fixed in addition.
The scaling relations for the critical exponents for $\eta=0$ \textcolor{black}{are}
\begin{equation}
 \alpha=2-3\nu\ , \ \beta=\nu/2\ ,\quad {\rm and}\quad  \gamma=2\nu\ .
\label{eqn:scale}\end{equation}

\begin{figure}
\includegraphics[width=12cm]{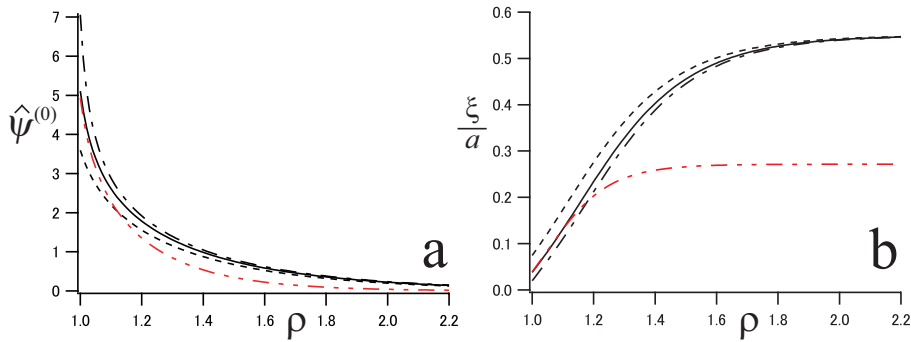}
\caption{The normalized order parameter (a) and correlation length (b) are plotted
\textcolor{black}{against the dimensionless radial distance $\rho\textcolor{black}{\equiv r/a}$.}
Dash, solid, and dot-dash curve represent the results for $\xi_\infty/a=0.55$, and respectively for
${\hat h}=24$, $60$, and $150$.  Two-dot chain (red) curve represents the result for  $\xi_\infty/a=0.27$ and
${\hat h}=60$.}
\label{static}
\end{figure}
Figure \ref{static} shows how the equilibrium order-parameter profile and the local correlation length depend on $\rho=r/a$.
Far from the particle, the former approaches zero, which represents the critical composition, and
the latter approaches the prescribed value of $\xi_\infty/a$.
Near the particle surface, the preferred component is more concentrated and
the correlation length becomes smaller as ${\hat h}$ increases. The stronger adsorption makes the mixture near
the surface more off-critical.
At a distance of $\xi_{\infty}$ from the particle surface,
$\psi^{(0)}$ is approximately reduced to $\psi_{\rm a}$.
The local correlation length $\xi$ should be smaller than the local length scale that the flow changes, denoted here by $l$,  
for the validity of the hydrodynamics formulated from the coarse-grained free-energy density.  In the flow 
\textcolor{black}{at a distance $r$ from the center of a particle}
moving translationally, $l$ would be equal or larger than \textcolor{black}{$r$}.  In Fig.~\ref{static}(b),
$\xi$  \textcolor{black}{is significantly small as compared with} $r$.  Thus, 
the hydrodynamics in the present formulation is available
in calculating the drag coefficient
for the parameter values examined.

It is known that the Gaussian free-energy density can describe the static properties only when the mixture is not so much
close to the critical point. 
The proportionality of $\xi\propto \tau^{-\nu}$ with
$\nu\approx 0.627$
is observed in the bulk of a binary fluid mixture
with the critical composition when it is sufficiently close to the critical point. The Gaussian free-energy density, giving 
$\xi\propto \tau^{-0.5}$ instead, appears to be valid when $\xi$
is smaller than approximately $15$ nm in the bulk of the mixture of 2, 6-lutidine and water \citep{jungk}.  
When the particle motion is studied in the Gaussian model \citep{Okamoto-Fujitani-Komura, Fujitani2018},
the correlation length is assumed to be homogeneous and much smaller than the minimum of $l$, {\it i.e.\/}, approximately $a$, 
for the validity of the hydrodynamics based on 
the free-energy density.
Thus, the Gaussian model is valid for small $\xi_\infty$ ($< 15$ nm in
the example above) and $\xi_\infty/a \ll 1$.  
The present formulation is not tied to these constraints.

\begin{figure}
\includegraphics[width=12cm]{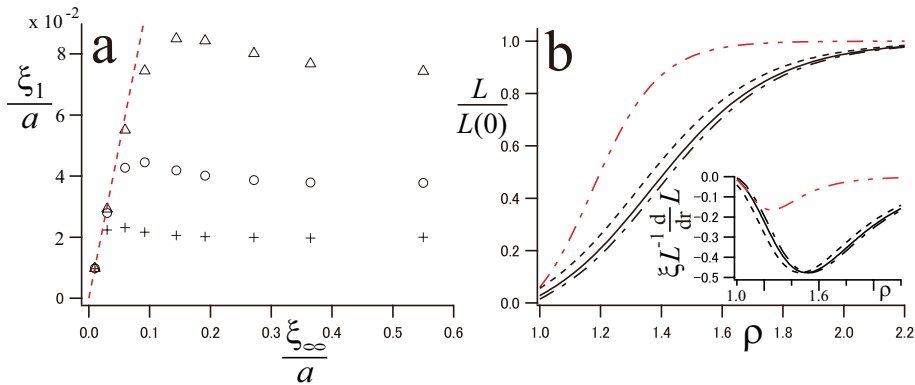}
\caption{ (a) The normalized correlation length at the particle surface $\xi_1/a$ is plotted
against  $\xi_\infty/a$ for ${\hat h}=24\ (\triangle),\ 60\ (\circ)$, and $150\ (+)$.  
Dash line (red) represents $\xi_1=\xi_\infty$.
(b) The ratio $L\textcolor{black}{(\psi^{(0)}(\rho a))}/L(0)$ is plotted against $\rho$. \textcolor{black}{
The normalized change of $L\textcolor{black}{(\psi^{(0)}(r))}$ over $\xi$, represented by
$\xi L^{-1} dL/(dr)$, is plotted against
$\rho\textcolor{black}{\equiv r/a}$ in the inset.}  The parameter values for each curve \textcolor{black}{are} the same as
used in Fig.~\ref{static}. }
\label{xixiLL}
\end{figure}

In Fig.~\ref{xixiLL}(a), $\xi$ at the surface (denoted by $\xi_1$) 
is approximately equal to $\xi_\infty$ for small $\xi_\infty/a$,
and reaches a plateau after a slight peak as $\xi_\infty/a$ increases.  
The plateau of $\xi_1$ \textcolor{black}{indicates that} ${\hat \psi}^{(0)}$ at the surface becomes independent of $\xi_\infty/a$.
The discrepancy between $\xi_1$ 
and $\xi_\infty$ implies the inhomogeneity in $\xi$, and cannot be described by  the Gaussian model.   As ${\hat h}$ increases in Fig.~\ref{xixiLL}(a),
the inhomogeneity appears \textcolor{black}{at} smaller $\xi_\infty/a$ and the plateau value is smaller, which is 
expected because the preferential  \textcolor{black}{adsorption} causes the inhomogeneity.  At the critical point ($\xi_\infty= \infty$),
the equilibrium profile around a spherical particle is calculated in \citep{Yabu-On}, where
$\xi_1/a$ is found to be given by $6^{-1/3}{\hat h}^{-2/3}$ for ${\hat h}\gg 1$. 
This theoretical result respectively gives $\xi_1/a=6.6, 3.6,$ and $1.9 \times 10^{-2}$  for ${\hat h}=24, 60,$ and $150$.
The values \textcolor{black}{of $\xi_1/a$} on the extreme right in Fig.~\ref{xixiLL}(a) are $7.4, 3.8$, and $2.0\times 10^{-2}$ for these values of ${\hat h}$,
respectively.  The latter two agree well with
the corresponding theoretical results, showing the strong adsorption ($\tau\ll \omega$ at the surface).
 Depending on the values of $a$ and ${\hat h}$, it is possible that the inhomogeneity occurs to invalidate
the Gaussian model even when the
Gaussian free-energy density gives good approximation to static properties in the bulk. 
Judging from Fig.~\ref{xixiLL}(a), when $a$ equals $100$ nm, the inhomogeneity begins to
occur at the value of $\xi_\infty$ smaller than $10$ nm 
for the strong adsorption.  At this small value of $\xi_\infty$, the Gaussian free-energy
density can describe the correlation length in the bulk of the mixture of 2, 6-lutidine and water,
\textcolor{black}{which is mentioned in the preceding paragraph} \textcolor{black}{and is also 
discussed in Sect.~\ref{sec:discuss}.}

The Onsager coefficient $L(\textcolor{black}{\psi^{(0)}})$ 
\textcolor{black}{increases as $\psi^{(0)}$ approaches zero, which} is shown in Fig.~\ref{xixiLL}(b).
At a distance of $\xi_{\infty}$ from the particle surface,
$L\textcolor{black}{(\psi^{(0)}(r))}$ increases to become approximately $70\ $\% of its value far from the particle, $L(0)$. Judging from
Figs.~\ref{static}(a) and \ref{xixiLL}(b), as is known, $\xi_\infty$ gives
a typical thickness of the adsorption layer.  
The inset of Fig.~\ref{xixiLL}(b) shows 
$\xi L^{-1} dL/(dr)$, whose absolute value represents \textcolor{black}{how large the variation} \textcolor{black}{of $L(\psi)$
over $\xi$,} \textcolor{black}{
mentioned below (\ref{eqn:Ldef}), is as compared with the} \textcolor{black}{local} \textcolor{black}{value of $L(\psi)$.}   
The absolute value is sufficiently small 
 for ${\hat h}=60$ over the region of $\rho$ examined,
while it is still smaller than unity but becomes rather large around $\rho=1.5$ for ${\hat h}=150$.  
\section{Results\label{sec:res}}
\begin{figure}
\includegraphics[width=8cm]{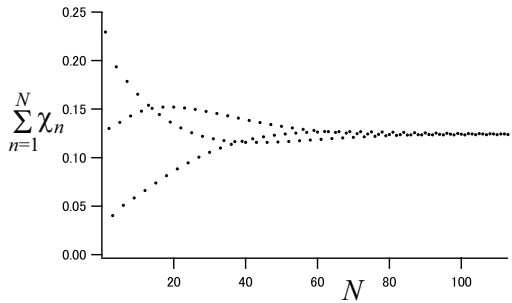}
\caption{ \textcolor{black}{Plot of $\chi_1+\chi_2+\cdots+\chi_N$ 
against $N$}
for ${\hat h}=150$ and $\xi_\infty/a=0.36$.}
\label{dgcoeff}
\end{figure}

We truncate \textcolor{black}{(\ref{eqn:chichi})} appropriately
to calculate $\Delta{\hat \gamma}_{\rm d}$.  The change of the partial sum occurring
when the number of the terms $(N)$ increases by one 
becomes smaller than $1\%$ of the partial sum for $N\ge 109$ in  
Fig.~\ref{dgcoeff}.  In this example, we can regard the partial sum of $N=109$
as the infinite sum.  As ${\hat h}$ or $\xi_\infty/a$ increases, we need larger $N$ 
to \textcolor{black}{evaluate} the infinite sum with the same accuracy, although data not shown. 
\textcolor{black}{The numerical results shown later are obtained using $N$ smaller than approximately $100$.  Some improvements to reduce the numerical cost are desirable when
$N$ is required to be significantly larger than $100$.}

\begin{figure}
\includegraphics[width=12cm]{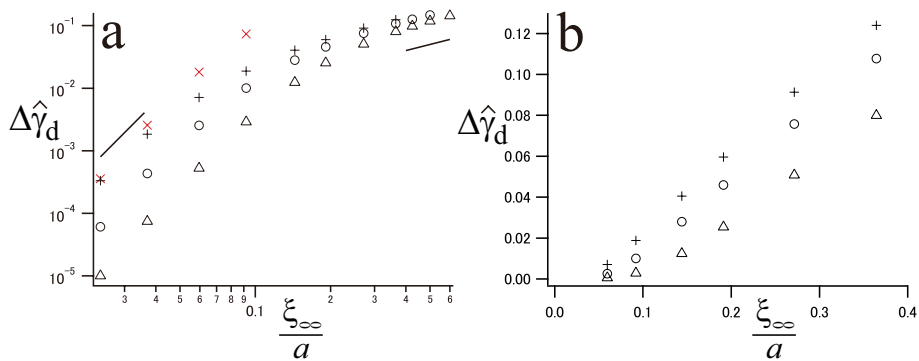}
\caption{(a) Logarithmic and (b) linear plots of \textcolor{black}{the normalized deviation of the drag coefficient}
$\Delta{\hat \gamma}_{\rm d}$ \textcolor{black}{against} $\xi_\infty/a$ for 
${\hat h}=24 (\triangle),\ 60 (\circ)$, and $150 (+)$. \textcolor{black}{Each of the results is obtained}
\textcolor{black}{using} \textcolor{black}{a partial sum of (\ref{eqn:chichi}).}
 Lines in (a) represent the slopes of one and four.
\textcolor{black}{The symbol $\times$ (red) in (a) represents the results in the Gaussian model for ${\hat h}=150$.} }
\label{dgvsxi}
\end{figure}

It is shown in Fig.~\ref{dgvsxi} how $\Delta{\hat \gamma}_{\rm d}$
changes as the critical temperature is approached.
The deviation increases with ${\hat h}$ and $\xi_\infty/a$, as is expected because it is caused by the 
adsorption.  In this figure, 
for small $\xi_{\infty}/a$, $\Delta{\hat \gamma}_{\rm d}$ appears to be 
\textcolor{black}{approximately} proportional to \textcolor{black}{the fourth power of $\xi_\infty/a$}.
As  $\xi_{\infty}/a$ increases in Fig.~\ref{dgvsxi}(a), the slope shown \textcolor{black}{by each of
the symbols, $+$, $\circ$, and $\triangle$,} becomes more gradual and the dependence 
appears to shift to the linear dependence;
the shift occurs at larger $\xi_{\infty}/a$ as ${\hat h}$ decreases.  
For large $\xi_\infty/a$ in Fig.~\ref{dgvsxi}(b), $\Delta{\hat\gamma}_{\rm d}$
appears to be a linear function of $\xi_{\infty}/a$, whose slope is 
calculated from the two data points on the right end  
to give $0.31$, $0.34$, and $0.35$ for ${\hat h}=24,\ 60$, $150$, respectively.
This suggests that the slope should become insensitive to ${\hat h}$ as $\xi_\infty/a$ increases.
The \textcolor{black}{approximate} fourth-power dependence for smaller $\xi_\infty/a$, \textcolor{black}{shown in Fig.~\ref{dgvsxi}(a),}
is mentioned  at the end of the penultimate paragraph of Sect.~4 of \citep{Fujitani2018}, where
the Gaussian model is studied \textcolor{black}{and the parameter $\Lambda(\approx {\hat h}/\sqrt{5\pi})$ is used.}
\textcolor{black}{For reference in Fig.~\ref{dgvsxi}(a),  
using the calculation procedure in \citep{Fujitani2018}, we plot the normalized deviation in the Gaussian model, corresponding 
with $\Delta{\hat \gamma}_{\rm d}$, for ${\hat h}=150$; the results ($\times$) 
deviate upward from \textcolor{black}{the results of the present study  ($+$) }
as $\xi_\infty/a$ increases.}

\begin{figure}
\includegraphics[width=12cm]{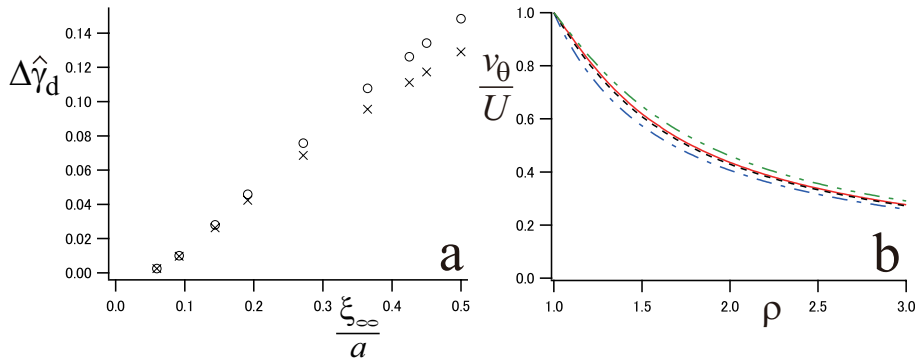}
\caption{(a) The normalized deviation $\Delta{\hat \gamma}_{\rm d}$ is obtained for ${\hat h}=60$
by assuming (\ref{eqn:Ldef}) ($\circ$)
and by assuming $L$ to be homogeneous ($\times$).  Some \textcolor{black}{of the}
\textcolor{black}{results shown by $\circ$ are shared with} Fig.~\ref{dgvsxi}. (b) 
The tangential component of the velocity field $v_\theta$ divided by $U$ 
is plotted against $\rho$ on the equatorial plane $\theta=\pi/2$.
Dot-dash curve (blue) represents the results in the absence of the preferential  \textcolor{black}{adsorption} ($h=0$).
Dash, solid, and two-dot chain curves (black, red, and green, respectively) 
represent the results for $({\hat h}, \xi_\infty/a)=(60, 0.19)$, $(150, 0.19)$, and $(60, 0.36)$, respectively. }
\label{const}
\end{figure}

By assuming $L$ to be homogeneous, instead of using (\ref{eqn:Ldef}),
we calculate $\Delta{\hat \gamma}_{\rm d}$ to obtain \textcolor{black}{the symbols of
$\times$} in Fig.~\ref{const}(a).
\textcolor{black}{This drastic change in modeling $L$} does not influence $\Delta{\hat \gamma}_{\rm d}$
for smaller values of $\xi_\infty/a$, and
reduces $\Delta{\hat \gamma}_{\rm d}$ \textcolor{black}{slightly}
as $\xi_\infty/a$ increases.  It is unchanged that
the dependence of $\Delta{\hat \gamma}_{\rm d}$ on $\xi_\infty/a$
becomes more gradual than in the Gaussian model. 
We thus expect that the appearance of the gradual dependence 
should be robust against the details of the dependence of $L$ on $\psi$,
although the results in the inset of Fig.~\ref{xixiLL}(b) suggest  that 
(\ref{eqn:Ldef}) is not completely reliable especially for ${\hat h}=150$.

As mentioned in Sect.~\ref{sec:intro}, some researchers regard the deviation of $\gamma_{\rm d}$
as caused by effective enlargement of the particle radius due to the adsorption layer.
However, for the parameter values examined in Fig.~\ref{const}(b),
the change of $v_{\theta}/U$ at $\theta=\pi/2$ due to the preferential  \textcolor{black}{adsorption} is smaller than approximately $20$ \%
of $v_\theta/U$ in its absence; the velocity field is not so much changed by the preferential  \textcolor{black}{adsorption} and
the mixture fluid in the adsorption layer cannot be regarded as a part of the rigid particle. The velocity gradient at the surface becomes more gradual
as $\xi_\infty/a$ increases, which suggests reduction in the viscous stress exerted on the particle.
The velocity field is influenced by $\textcolor{black}{\boldsymbol{\Pi}}$ because of (\ref{eqn:nabpsi}) \textcolor{black}{and} (\ref{modelH2}). 
How $\textcolor{black}{\boldsymbol{\Pi}}$ changes with $\xi_\infty/a$ \textcolor{black}{influences on how} 
$\Delta{\hat \gamma}_{\rm d}$ \textcolor{black}{changes with} $\xi_\infty/a$
not only directly through the first term in the parentheses of (\ref{eqn:dragintegral}) but also through the last two terms
by changing the flow field.  

The dependence of $\Delta{\hat \gamma}_{\rm d}$ on $\xi_\infty/a$
becomes close to the linear one in Fig.~\ref{dgvsxi}(a)
in the range of $({\hat h}, \xi_{\infty}/a)$ showing the inhomogeneity of $\xi$ in Fig.~\ref{xixiLL}(a).   
This is reasonable because of the close relationship between $\psi^{(0)}$ and $\Delta\gamma_{\rm d}$ shown by (\ref{eqn:draco}). 
\textcolor{black}{In Fig.~\ref{xixiLL}(a),} the inhomogeneity occurs approximately 
when $\xi_1/a$ exceeds $\xi_\infty/a$. Thus, when large ${\hat h}$ causes the strong adsorption, the dependence in Fig.~\ref{dgvsxi}(a)
becomes close to the linear one if we have
\begin{equation}
\frac{\xi_\infty}{a} > 6^{-1/3}{\hat h}^{-2/3} \label{eqn:estima}
\ .\end{equation}
\section{Discussion\label{sec:discuss}}
\textcolor{black}{The Gaussian free-energy density was used in the previous studies on the drag coefficient 
\citep{Okamoto-Fujitani-Komura, Fujitani2018}.  
There, $\xi_\infty$ should be sufficiently small so that the free-energy density can describe the static properties in the bulk
and so that the correlation length can be regarded as homogeneous.  The latter condition can require
smaller $\xi_\infty$  than the former \textcolor{black}{unless the adsorption is weak}. 
Thus, the Gaussian model
is not always available in calculating $\gamma_{\rm d}$ \textcolor{black}{even if the static properties in the mixture bulk}
can be described by the Gaussian free-energy density, as mentioned in \textcolor{black}{Sect.~\ref{sec:nond}.}
The free-energy density in (\ref{eqn:largeF}) is completely free from these conditions
because it can describe the static properties in the bulk even at the critical point
and the inhomogeneity of $\xi$.  This inhomogeneity is correlated with the appearance of the approximate
linearity between $\Delta{\hat \gamma}_{\rm d}$ and $\xi_\infty/a$, which should occur for large $\xi_\infty/a$ satisfying 
(\ref{eqn:estima}). } 

\begin{figure}
\includegraphics[width=6cm]{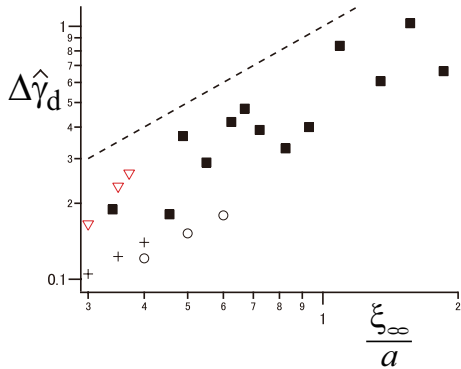}
\caption{Solid squares represent part of the data in the top of Fig.~2 of \citep{Omari}.
Circles and crosses represent \textcolor{black}{our
results calculated} with $\eta=0.036$ for ${\hat h}=60$ and $150$, respectively.
The calculated values are slightly larger for large $\xi_\infty/a$ than the ones obtained with $\eta=0$.
For $({\hat h}, \xi_\infty/a)=(60, 0.4)$ and $(60, 0.5)$, we have $\Delta{\hat \gamma}_{\rm d}=0.119\ (0.121)$ and $0.148\ (0.152)$  
by using $\eta=0\ (0.036)$, respectively. \textcolor{black}{The triangles (red) represent the results for ${\hat h}=24$ in the Gaussian model.
Dash line represents $\Delta{\hat \gamma}_{\rm d}=\xi_\infty/a$.} }
\label{omari}
\end{figure}

Omari {\it et al\/} \citep{Omari} measured the self-diffusion of silica particles in a near-critical mixture of 2,6-lutidine and water, 
for which
$T_{\rm c}$ is approximately $307\ $K \citep{gulari, jungk}.
The former is preferentially \textcolor{black}{adsorbed} by the particle, and the self-diffusion coefficient is suppressed as $T_{\rm c}$ is approached.
The data on the top of Fig.~2 of \citep{Omari}, except the data
point on the extreme left, are replotted in Fig.~\ref{omari};
$a=25\ $nm and $\xi_0=0.25\ $nm used in this reference lead to $\tau_{\rm a}=6.5\times 10^{-4}$.
We thus have $\xi_\infty/a=0.76$ for $\tau=10^{-3}$; the replotted data range from $\tau=3.6\times 10^{-3}$ to $2.3\times 10^{-4}$.  
Equations (\ref{eqn:nodim}) and (\ref{eqn:hath})
give ${\hat h}=h \sqrt{ 3u^*a^3}/\left(k_{\rm B}T_{\rm c}\sqrt{C_1}\right)$ for $\eta=0$. 
The value of $h$ may be usually smaller than $10^{-5}$ m$^3/$s$^2$,
considering the following three points; (1) a typical energy of the hydrogen bond is $10^{-20}$ J per a molecule,
(2) the area of the particle surface interacting with one molecule of the mixture
would be larger than approximately $1$ nm$^2$, and (3) $|\psi|$ is at most $10^3$ kg/m$^{3}$.
 Using a typical value of $M$ for alkanes, $10^{-16}\ $m$^7/($kg$\cdot$ s$^2)$ 
\citep{Mref1,Mref2}, we find that ${\hat h}$ is smaller than approximately $160$.  \textcolor{black}{
We use a partial sum of (\ref{eqn:chichi}) with $\eta=0.036$
to calculate $\Delta{\hat \gamma}_{\rm d}$, and plot the results for small values
of $\xi_\infty/a$ in Fig.~\ref{omari}.}  \textcolor{black}{We did not calculate $\Delta{\hat \gamma}_{\rm d}$
at values of $\xi_\infty/a$ larger than shown in the figure for each of ${\hat h}=60$ and $150$
because of heavy numerical costs.}
The calculation results for $h=60$ and $150$ appear to have almost the same slopes with the slope suggested by the experimental data. 
However, judging from Fig.~\ref{omari}, the calculation would 
yield smaller $\Delta{\hat \gamma}_{\rm d}$ than the experimental
data, even if performed for the same values of $\xi_\infty/a$
that gives the experimental data.  The underestimation may be attributed to our assumption 
of  homogeneous viscosity.

The singular part of the viscosity is proportional to $\xi^{1/19}$ up to the 1-loop order
\citep{ohta, ohtakawasaki, sigg}.  Thus, its nonuniversal regular part, which can depend on the local composition,
 is not negligible unless the mixture is very close to the
critical point.  We can estimate $\psi_{\rm a}\approx 15\ $kg/m$^3$ for the parameter values mentioned above, using  
(\ref{eqn:defC2}), (\ref{eqn:Mdef}) and (\ref{eqn:nodim}).
Considering that mass density of 2, 6-lutidine ($0.93\ $g/cm$^3$) is almost the same as that of water,
the change of $5\%$ in the weight percent of 2, 6-lutidine  \textcolor{black}{in the mixture} approximately amounts \textcolor{black}{to} that of $7$ in ${\hat\psi}^{(0)}$.
In Fig.~\ref{static}(a), the values of ${\hat\psi}^{(0)}$ at the surface are approximately $5$ and $7$ for ${\hat h}=60$ and $150$ \textcolor{black}{for $\xi_\infty/a=0.55$}, respectively. 
The data of the viscosity for various values of the weight percent are shown at $\tau\approx 10^{-2}$ and 
$10^{-4}$ in \citep{Stein}.   From them, in the experimental system yielding the data
replotted in Fig.~\ref{omari},   
the viscosity near the surface is guessed to be raised by approximately $30\ \%$
around $\tau=10^{-3}$. \textcolor{black}{It is thus possible that the underestimation suggested in
Fig.~\ref{omari} can be explained by this slight increase of
the viscosity in the adsorption layer. The viscosity
may be regarded as homogeneous over the mixture around $\tau=10^{-4}$ due to the
overlap of the regular and singular parts,  
according to the data in \citep{Stein}}.  \textcolor{black}{These data} 
\textcolor{black}{also suggest that, when $\tau$ is smaller than $10^{-4}$, or $\xi_\infty$ is
larger than approximately $80$ nm, 
the singular part should become more explicit to 
cause inhomogeneous viscosity.}

\textcolor{black}{The results 
in the Gaussian model \textcolor{black}{for ${\hat h}=150$} in Fig.~\ref{dgvsxi}(a) 
are obtained using a series expansion with respect to the dimensionless parameter
proportional to $h$, which is mentioned in the preface of Sect.~\ref{sec:calc}}\textcolor{black}{;  the} 
\textcolor{black}{series is given by
the first entry of (47) in \citep{Fujitani2018}.} \textcolor{black}{
For ${\hat h}=24, 60$ and $150$, respectively,
when $\xi_\infty/a$ is larger than approximately $0.4$, $0.18$ and $0.1$,}
\textcolor{black}{the drag coefficient} \textcolor{black}{cannot be calculated 
in the Gaussian model because the expansion series
is numerically divergent.} 
\textcolor{black}{Results in the Gaussian model  for ${\hat h}=24$ 
are plotted with triangles in Fig.~\ref{omari}.}
\textcolor{black}{
For ${\hat h}=60$ and $\xi_\infty/a=0.14$,} \textcolor{black}{the normalized deviation in} \textcolor{black}{the Gaussian model is
$8.3\times 10^{-2}$,} \textcolor{black}{and is
larger than the corresponding result shown by the circle}  \textcolor{black}{
in Fig.~\ref{dgvsxi}(a).
Thus, as far as examined, the dependence of the normalized deviation of the drag coefficient on $\xi_\infty/a$ in the Gaussian model
does not become close to the linear one, in contrast to} \textcolor{black}{our results} \textcolor{black}{calculated
with} \textcolor{black}{the} \textcolor{black}{renormalized local functional theory.}
\textcolor{black}{It is to be noted that, judging from Fig.~\ref{xixiLL}(a),
the Gaussian model loses validity for ${\hat h}=24, 60,$ and $150$ in the range of $\xi_\infty/a$ of Fig.~\ref{omari}. 
It is thus only for reference that the results in the Gaussian model are plotted in this figure. }
\textcolor{black}{
If the particle were enlarged by the thickness of the adsorption layer, we would have $\Delta{\hat \gamma}_{\rm d}=\xi_\infty/a$,
which is also plotted for reference in Fig.~\ref{omari}.} \textcolor{black}{This naive idea clearly oversimplifies the dynamics of the mixture;
it is suggested by Fig.~\ref{const}(b) that the mixture generally flows in the adsorption layer. 
That the dash line lies above the experimental data in Fig.~\ref{omari}
may imply that the naive idea could be justified approximately if the preferred component were extremely viscous.}

The particle radius in the experiment is rather small.  For $a=25\ $nm, 
our coarse-grained picture should become less reliable very near the surface because
$\xi_1$ reaches a microscopic length scale in Fig.~\ref{static}(b).
Our theory can be safely applied to the particle radius larger than approximately $100\ $nm.
Lee \citep{Lee} measured the self-diffusion coefficient of such large particles in a ternary mixture
near the plait point, and found out the linear dependence of $\Delta{\hat \gamma}_{\rm d}$ on $\xi_\infty/a$
when \textcolor{black}{$\tau$} is approximately smaller than $3\times 10^{-2}$ ({\it i.e.\/}, $\xi_\infty\stackrel{>}{\sim}5\ $nm).  
For this mixture, the renormalized local functional theory is available if $\tau$ is replaced by a variable
proportional to $\tau^{1/(1-\alpha)}$ \citep{fisher,folk}, although
\textcolor{black}{the definition of the surface field} and  (\ref{modelH1})  
should be modified for ternary mixtures.

\textcolor{black}{In} Fig.~\ref{static}(b),
$\xi$ is much smaller than $a$ near the particle, and increases
to approach $\xi_\infty$ mainly in the region of $r-a<\xi_\infty$.  \textcolor{black}{Thus, for the parameter values} \textcolor{black}{
examined,} $\xi$ nowhere exceeds  
the length scale that the flow changes \textcolor{black}{($l$)}, which 
would be equal or larger than $r$.  \textcolor{black}{In particular when the adsorption strong, the
local correlation length near the particle is much smaller than the particle radius.
Thus,} when we consider a large
particle ($a\stackrel{>}{\sim} 100\ $nm) \textcolor{black}{with the strong adsorption} in a near-critical mixture 
\textcolor{black}{whose $\xi_\infty$ is not so large as to cause explicit singular part of the viscosity}
($\xi_\infty<80\ $nm
in the mixture mentioned above), $l$ is sufficiently large as compared with
$\xi$ everywhere.  \textcolor{black}{Then, we can use the present
hydrodynamic formulation based on the coarse-grained free-energy
density in calculating the drag coefficient, without considering the critical fluctuations
significant on length scales smaller than $\xi$.}

We \textcolor{black}{use  $\varepsilon$} as a convenient parameter {\textcolor{black}{to
calculate the drag coefficient up to the linear order}
with respect to the particle speed.  The parameter is taken as representing the Reynolds number
in deriving the conventional Stokes law, which holds for the low Reynolds number, as mentioned in Sect.~\ref{sec:prel}.
Thus, we can discuss the validity range of the linear regime
by making the physical meaning of $\varepsilon$ explicit.
In the presence of the preferential adsorption in the near criticality,
the low Reynolds number should remain required for the linearity, but it would not be
sufficient  because nonlinearity is suggested to be \textcolor{black}{made relevant} by
significant distortion of the adsorption layer due to the particle motion \textcolor{black}{\citep{furu}.
The distortion would be reduced by the mutual diffusion of the two fluid components, while
augmented by the convection. 
The time scale that the former occurs over the adsorption layer can be estimated to be $(a+\xi_\infty)^2/D$, 
where $D$ is the diffusion coefficient given by the fraction in (\ref{eqn:mct}),
while the one for the latter to be $a+\xi_\infty$ divided by the particle speed $U_{\rm o}$.  
Thus, $\varepsilon$ in our problem would be taken as representing not only the Reynolds number but also
the P{\' e}clet number, $(a+\xi_\infty)U_{\rm o}/D$, which increases with $\xi_\infty$.
It is to be noted that we calculate the drag force by taking into account the distortion of the adsorption layer,
as mentioned above (\ref{eqn:piforce}).  However, elucidating the validity range should require calculating beyond the linear regime
and checking the linearity for various values of the parameters including $\xi_{\infty}$.
 To examine whether the Brownian motion 
can be described by a linear Langevin equation,
we may need to study the drag force beyond the linear regime in
the framework of the fluctuating hydrodynamics, considering that a Brownian particle does not always move 
translationally in a quiescent fluid. 
It is thus well expected that the self-diffusion coefficient of a Brownian particle should not always follow
Einstein's relation, {\it i.e.\/}, should not be always given by $k_{\rm B}T/\gamma_{\rm d}$, even when the
Reynolds number is sufficiently \textcolor{black}{small. The relation can break down} when $\xi_\infty/a$ is sufficiently large.
We believe that our results of the drag coefficient should} \textcolor{black}{also} \textcolor{black}{give a firm basis to some
future works, experimental or theoretical, on the validity range of $\xi_\infty/a$ for
Einstein's relation in the presence of the preferential adsorption in the near-criticality.  }

\section*{Acknowledgements}
S. Y. was supported by Grant-in-Aid for Young Scientists (B) (15K17737 and 18K13516).

\appendix
\begin{section}{Some details}
The kernel appearing in (\ref{inteq-R}) is given by
\begin{equation}
\varGamma_{R}\left(\rho,\sigma\right)=\begin{cases}
\frac{\left(3-5\sigma^{2}\right)}{2\sigma^{3}\rho^{3}}+5\frac{\left(3\sigma^{2}-1\right)}{2\sigma^{3}\rho}+\sigma^{2}\rho^{-3}-5\rho^{-1} & \rho\geq\sigma\\
\frac{\left(3-5\sigma^{2}\right)}{2\sigma^{3}\rho^{3}}+5\frac{\left(3\sigma^{2}-1\right)}{2\sigma^{3}\rho}+\rho^{2}\sigma^{-3}-5\sigma^{-1} & \sigma\geq\rho.
\end{cases}
\end{equation}
We find
\begin{equation}
\left( {1\over 2}\partial_\rho^3+2\partial_\rho^2\right) \varGamma_{\rm R}(\rho, \sigma) 
=15\left[\alpha_0(\sigma)+1\right]\ ,
\label{eqn:alas}\end{equation}
\textcolor{black}{\textcolor{black}{at $\rho=1$.} The operator in the parentheses above appears in (\ref{eqn:onecompdragz}).}
The kernel appearing in (\ref{inteq-Q}) is given by
\begin{align}
\varGamma_{Q}\left(\textcolor{black}{\rho},\sigma\right) & =\begin{cases}
\left(\textcolor{black}{\rho} \sigma\right)^{-2}/2+\textcolor{black}{\rho}\sigma^{-2} & \textcolor{black}{\rho}\leq\sigma\\
\left(\textcolor{black}{\rho}\sigma\right)^{-2}/2+\sigma \textcolor{black}{\rho}^{-2} & \sigma\leq \textcolor{black}{\rho}.
\end{cases}
\end{align}
These kernels are originally obtained in \citep{Okamoto-Fujitani-Komura}.

We can calculate the equilibrium profile numerically by using (\ref{eqn:mu0}) and (\ref{eqn:psiatsurface}),
as they are, after the non-dimensionalization mentioned in Sect.~\ref{sec:nond}.
Alternatively, we can utilize (A5) of \citep{rigid}, whose $s\left(\psi(\rho)\right)$ is defined by $\omega(\psi(\rho))/\tau-1$ 
becomes proportional to
$e^{-2\rho a/\xi_\infty}/\rho^2$ far from the particle.  The differential equation with respect to
the ratio of the former to the latter, that is $s (\psi(\rho)) e^{2\rho a/\xi_\infty} \rho^2$, can be derived from (\ref{eqn:mu0}),
and is solved numerically under the boundary conditions imposed sufficiently far from the particle.
At $\rho =10$, we required the derivative of $s$ with respect to $\rho$ to vanish, and 
fixed the value of $s$ so that ${\hat h}$ becomes a prescribed value
with the aid of (\ref{eqn:psiatsurface}).
\end{section}

\bibliographystyle{jfm}
\bibliography{jfm-manuscript-resub825}

\end{document}